## MATHEMATICAL MODELS AND COMPUTATIONAL METHODS

# Using Reinforcement Learning in the Algorithmic Trading Problem

E. S. Ponomarev[a, *], I. V. Oseledets[a, b], and A. S. Cichocki[a]

[a]*Skolkovo Institute of Science and Technology, Moscow, Russia*
[b]*Marchuk Institute of Numerical Mathematics, Russian Academy of Sciences, Moscow, Russia*
*\*e-mail: Evgenii.Ponomarev@skoltech.ru*

**Abstract**—The development of reinforced learning methods has extended application to many areas including algorithmic trading. In this paper trading on the stock exchange is interpreted into a game with a Markov property consisting of states, actions, and rewards. A system for trading the fixed volume of a financial instrument is proposed and experimentally tested; this is based on the asynchronous advantage actor-critic method with the use of several neural network architectures. The application of recurrent layers in this approach is investigated. The experiments were performed on real anonymized data. The best architecture demonstrated a trading strategy for the RTS Index futures (MOEX:RTSI) with a profitability of 66% per annum accounting for commission. The project source code is available via the following link: http://github.com/evgps/a3c_trading.

## 1. INTRODUCTION

The algorithmic trading considered in this paper consists in designing a control system capable of buying or selling a fixed volume of a financial instrument on the stock exchange. The algorithm is intended to maximize the cost of the total portfolio or, in other words, the profit. As the financial instrument, we considered RTS Index futures in our project; the data for the experimental part were obtained from a large Russian exchange. Commission is taken into account according to the prices of this exchange for futures trading.

The design of the algorithm is based on using reinforcement learning since this is most appropriate for problems with delayed reward. In contrast to supervised learning, this does not require creating the rules under which a certain action must be considered true with a certain weight and allows using the metrics calculated for each strategy for long time intervals, for instance, the Sharpe ratio [4]. We use a modified asynchronous advantage actor-critic algorithm [12] in our work. As the approximation function, we studied several artificial neural network (ANN) architectures. We show the dependence of the results on a variation in network depth and number of parameters (neurons) and on the introduction of a recurrent layer, namely, long short-term memory (LSTM) [2]. Data in the form of anonymized bids were aggregated empirically in the vector of attributes. This action is chosen once per sixty seconds.

Note that applicability of all developed methods was confirmed experimentally on real data. We detail the principal results in the next sections.

Today, there are multiple reinforcement learning algorithms [5] and parts of them have been applied in algorithmic trading, for instance, in Q-learning [6], Deep Q-learning [1, 7], recurrent reinforcement learning, and policy gradient methods [8, 6, 9], REINFORCE [10], and other actor-critic methods [5, 11]. However, this research area is rapidly developing and new algorithms appear.

In this work we construct an environment [3] typical for a reinforcement learning problem, which consists of **states**, a set of possible **actions**, and a **reward** function. In addition, we assume that this process has the Markov property. As the learning method we applied the **asynchronous advantage actor-critic** (A3C) algorithm [12], which showed efficiency on many datasets, including Atari 2600 [13]. The application of this algorithm to the exchange trading problem was not found in the literature. In the improvement process of the algorithm operation, we searched for artificial neuron network architecture lying in the basis of the method, including the study of recurrent neuron networks and several other optimizations. The algorithm was learned and tested on real anonymized data on the trading of the RTS Index at the Moscow Exchange (MOEX:RTSI).

Notable results of the work:

(1) We studied the application of the reinforcement learning algorithm based on the deep neuron network.

(2) The modern asynchronous advantage actor-critic (A3C) algorithm was applied to these tasks.

(3) We searched for artificial neuron network architecture.

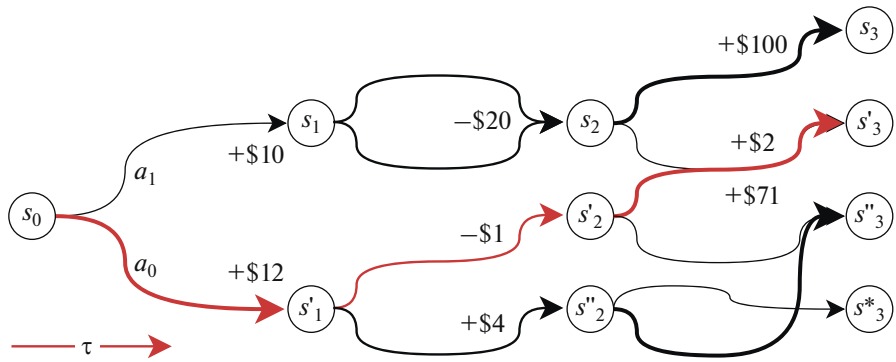

**Fig. 1.** τ is the red track, *a* is the action, *s* is the state, and *r* is the reward.

(4) The different architectures of the method were experimentally probed and compared. The best demonstrated a stable winning strategy with a profitability of 66% per annum over six months of testing (accounting for exchange commission).

## 2. PRELIMINARY INFORMATION

### 2.1. Formulation of the Exchange Trading Problem in Terms of Reinforcement Learning

Consider the exchange trading process in more detail. Let us fix a financial instrument (futures on the RTS index) and its possible volume and let us agree to make a decision on buy/sell once per time quantum, the minute. Thus, we obtain the system where in each step we need to choose one of three actions, the desired position. The position may be either long, when we possess the futures; neutral, when we turn all assets into money; or short, when we borrow a fixed volume under the obligation to buy it later at the future price. In addition, new data become available at each step, which may be aggregated in the form of a state vector.

We prescribe the environment as the Markov decision-making process [14]: $M = \{S, A, P, \gamma, R\}$, where

(i) $\mathbb{S} \in \mathbb{R}^m$ is the space of **observed states**. At each step the exchange-agent system is in $s_t \in S$. $s_t$ is constructively represented in the work as an aggregation of current bids or as an internal state of the LSTM memory cell. In the last case, we may hope to procure the larger informative value.

(ii) $\mathbb{A} = \{-1, 0, 1\}$ is the space of **actions**. In the trading problem, the action is the desired position: long (to keep a unit of the instrument, long, 1), neutral (cash out, 0), or short (borrow a unit of the instrument, short, −1).

At each step we choose the action $a_t \in A$ from the developed **politics** $\pi(a|s)$, that is, the probability to choose $a$ in $s$.

(iii) $P(s'|s, a)$ is the **transition probability** of the assumed Markov process.

(iv) $R(s, a)$ is the **reward** function. At each step the agent becomes the reward depending, not only on the current, but also on the previous actions $r_t = R(s_t, a_t)$.

(v) $\gamma \in [0, 1]$ is the decay multiplier with which the next reward is summed up into the total reward for an action $R_t = \sum_{i=0}^{T} \gamma^i r_{t+i}$.

**Statement 1. The task of the algorithm** *is to find the strategy* $\pi: \mathbb{S} \to \mathbb{A}$ *that maximizes the mathematical expectation of reward* $\rho^\pi$:

$$\rho^\pi = \rho^\pi(\theta) = \mathbb{E}[R|\pi] \to \max_\pi,$$

$$\rho^\pi = \mathbb{E}[R|\pi(\theta)] = \int_T p(\tau|\pi(\theta)) R(\tau) d\tau \to \max_\theta,$$

*where the track* $\tau = \{s_t, a_t\}_{t=0}^T \in T$ *is the realization of one game/episode (Fig. 1) and* $R(\tau)$ *is the total reward.*

As the reward we use its estimate, the function $R(s, a)$ of action $a$ and state $s$

$$R(s, a) \approx V^\pi(s) + A^\pi(s, a),$$

where $A^\pi(s, a)$ is the advantage function, which is the value characterizing how much chosen action $a$ is better than the average estimate of the utility value of state $s$

$$A^\pi(s, a) := R(s, a) - V^\pi(s)$$

and $V^\pi(s)$ is the estimate of the cost function of state $s$ for the politics $\pi: \mathbb{S} \to \mathbb{A}$:

$$V^\pi(s) = \mathbb{E}_\pi R^\pi(s, a).$$

Thus, the architecture of the actor determines which action $a = \pi(s)$ to choose. The loss function for this subnetwork is in linear proportion to advantage function $A(s_t, a_t) = [r_t + \gamma V^\pi(s_{t+1})] - V(s)$. The gradient of the loss function for this part of the network will equal (the derivation may be found in [12])

Actor: $\nabla_\theta J(\theta) \approx \mathbb{E}_\pi[\nabla_\theta \log \pi_\theta(a|s) A^\pi(s, a)]$.

The architecture of the critic predicts which reward is anticipated in this state without dividing the esti-mates into actions $V^\pi(s_t) = E_\pi R_t(\pi)$. The loss function for the critic is the error in the prediction $V_\pi(s_t)$:

$$\text{Critic: } L_v = \sum_{i=1}^{T}(V_{\text{target}_i} - V_i)^2,$$

$$V_{\text{target}_i} = \sum_{k=i}^{T}\gamma^{k-i} r_i.$$

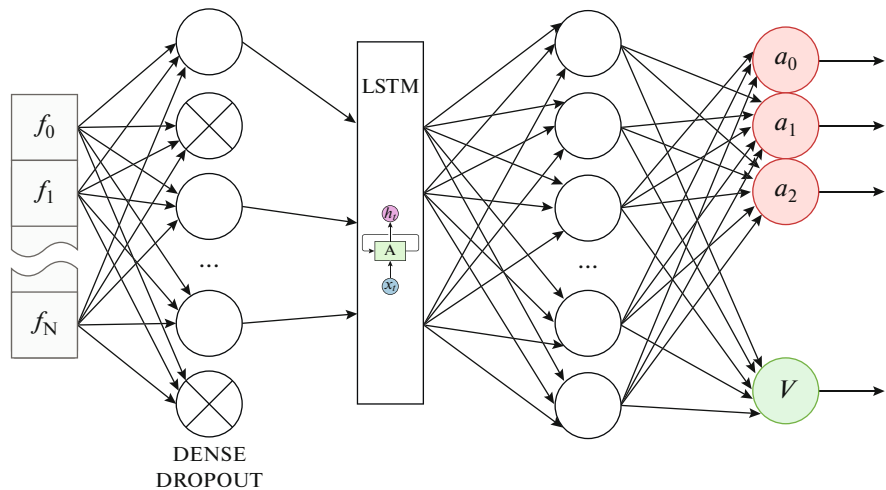

**Fig. 2.** Artificial neuron network determining cost function $V(s) = V(s|\theta)$ and politics (actor) function $\pi(a|s) = \pi(a|s, \theta)$.

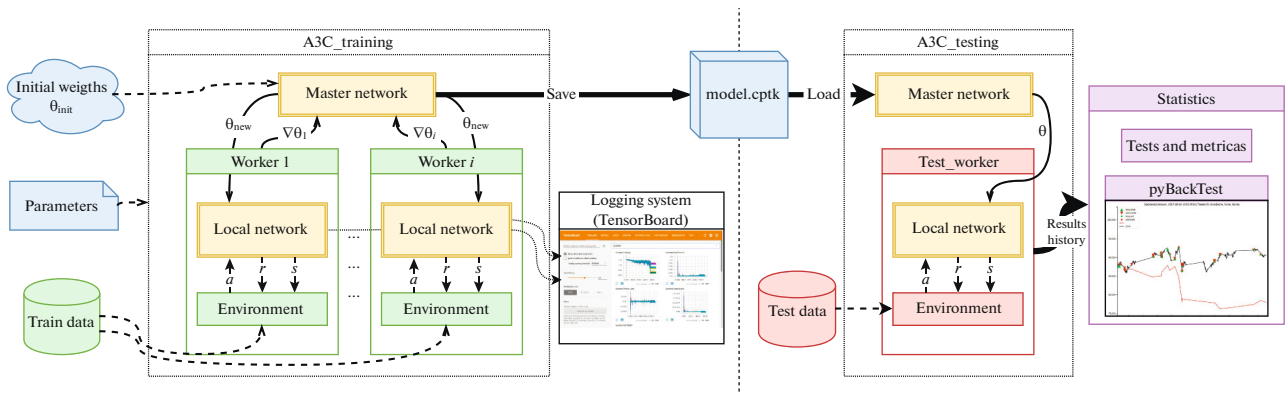

**Fig. 3.** Scheme of training and testing system.

In the problem considered, $P$ and $R$ are unknown, and the instantaneous reward function reflects variation in the portfolio cost and takes into account the commission from the buy/sell of an instrument. It means that, if $c_t$ is the cost of all assets of the agent at step $t$ (the amount of money from selling everything without regard for commission), then $r_t = c_t - c_{t-1} - fee \cdot \mathbb{I}[a_t = a_{t-1}]$.

The entire network is trained with the error-back-propagation method. The loss function for the entire network is common and is a linear combination of the loss functions for the actor and critic (parameter $\alpha \in [0, 1]$):

$$L = \alpha\sum_{i=1}^{T}(V_{\text{target}_i} - V_i)^2 - \log\pi_\theta(a_i|s_i)A_i^\pi. \quad (1)$$

As a result, the algorithm attempts to maximize the cumulative discounted reward

$$R_t = \sum_{i=0}^{T}\gamma^i r_{t+i},$$

where $r_{t+i}$ is a variation in the portfolio for a step. The number of steps is $T \sim 200$, because the majority of terms make no considerable contribution to the estimate ($\gamma^T \to 0$).

## 3. PROGRAM IMPLEMENTATION

The actor-critic algorithm execution with several asynchronous flows shows a significant improvement in the quality of operation [12]. For the purposes of study, we constructed the software system of training and testing. We implement the system using Python within the TensorFlow framework [18]. The system consists of two global processes: training and testing (Fig. 3). They are associated by storing the model in the hard disk and loading the architecture and the trained parameters in the testing system. The choice of architecture is a key task and is discussed in detail in Section 4.

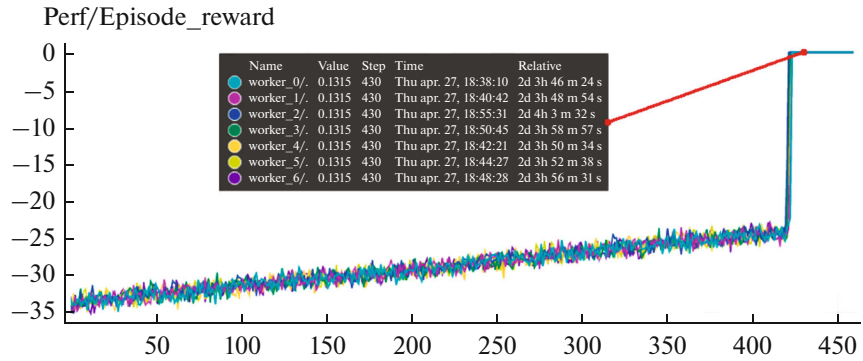

**Fig. 4.** Training curve of algorithm. Falling into the extremum corresponding to the buy-and-hold strategy is marked.

The training of the neuron network is performed every $N_{steps} = 200$ steps. $N_{worker}$ is the number of parallel processes, which was fixed in this work equal 10. Each epoch of training was conducted on the three-month data containing 50 000 one-minute steps. The convergence requires approximately 1000 epochs for a step size of approximately $10^{-3}$.

To accelerate the system testing, we implemented the testing subsystem without training and writing in the history. We replace the probabilistic approach in the testing with the arg max function. We also switch off the dropout function in the testing.

## 4. NUMERICAL RESULTS

We performed numerical experiments on real historical data (the complete log of the anonymized bids for the RTSI). As the training set we used the bids from September 15, 2015 to December 15, 2015. We carried out the test over the following six months: from December 15, 2015 to June 15, 2016.

The commission from buy or sell was fixed at 1.25 rubles for an operation, that is, 2.50 rubles for each transaction. It is clear that the strategy must be more profitable than 2.50 rubles for a trade. To increase the profitability of each transaction, we artificially increased the commission in the reward model. In addition, we changed the reward function to avoid coming into the buy-and-hold trap (Fig. 4), which means no active trading. To do this we introduced a penalty for the long repetition of an action.

The development of neural network architecture is important. As the initial point we chose the simplest artificial neural network with a single common hidden layer, linear function $V$, and a linear layer with softmax activation for choosing action $\pi(s)$. As the hypotheses improving method quality, we made the following assumptions:

**Assumption 1.** Using a different reward function.

**Assumption 2.** Introducing a recurrent layer (LSTM).

**Assumption 3.** Adding a dropout layer.

**Assumption 4.** Increasing the number of neurons in the hidden layers.

**Assumption 5.** Using a more complicated architecture of the cost function.

**Assumption 6.** Combining the attributes for several minutes in a common vector.

Following these assumptions, we designed several architectures from combinations of the same layers, but with different parameters, including no layer (Table 1 and Fig. 2). In Table 1 we used the following denotations:

(1) Depth is the number of serially connected vectors of attributes used as the input vector.

(2) Dense is the number of neurons in the fully connected first layer (for instance, 128) or the absence of this layer (–).

(3) Dropout is the probability of dropout (for instance, 0.5) or its absence (–).

(4) LSTM is the number of neurons in the LSTM layer connected with the first layer (for instance, 64) or the absence of this layer (–).

(5) Dense V is the number of neurons in the fully connected layer preceding the output linear critic layer ($V(s)$).

**Table 1.** Architectures

| Name | Depth | Dense | Dropout | LSTM | Dense V | Dense A |
|---|---|---|---|---|---|---|
| 5 | 6 | – | 0.5 | 64 | – | – |
| 8 | 6 | – | 0.5 | 128 | – | – |
| 5coolV | 6 | – | 0.5 | 64 | 32 | – |
| 9 | 1 | – | 0.5 | 64 | 32 | – |
| 12 | 1 | – | 0.5 | 64 | 32 | 32 |
| 5noLSTM | 20 | – | – | – | – | – |
| 6 | 6 | 128 | – | 128 | – | – |

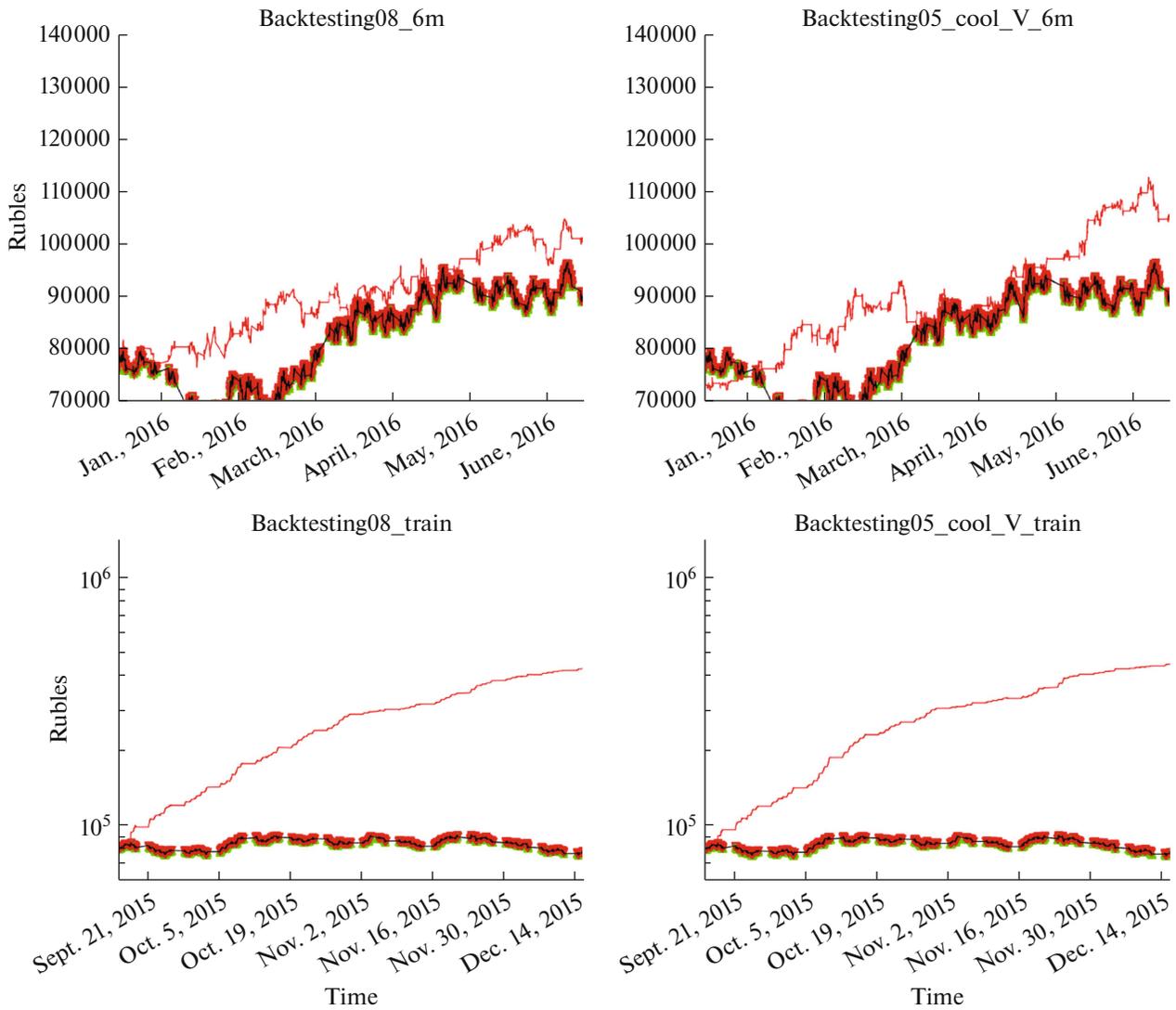

**Fig. 5.** Comparison of 8 and 5 *cool* V models: effect of linear approximation of cost function of state $V(s)$ (to the left) against a two-layer network with activation function tanh at first layer. The remaining parameters are identical (Table 1).

(6) Dense A is the number of neurons in the fully connected layer preceding the output softmax actor layer ($\pi(a|s)$).

To study how the results depend on the presence of the dropout layer, we chose the two architectures named 6 and 8. From the test results (Table 3) the dropout significantly improves the situation. To compare result dependence of the number of neurons in the LSTM layer, we considered 8 and 5 architectures, and, to compare the dependence on the complicatedness of the cost function approximator, we used 5 and 5 *cool* V architectures. To check how the results depend on the number of serially connected vectors of attributes used as the input vector, we considered 5 *cool* V and 9 architectures (the training and testing curves are depicted in Figs. 5–7).

**Table 2.** Result of execution on three-month test data

| Name | Profit % per annum | Profit % per annum (commiss.) | Sharpe ratio | Fraction of winning transactions | Average transaction, rubles |
|---|---|---|---|---|---|
| 5 | **99.0** | **94.9** | **3.23** | **60.32** | **59.59** |
| 8 | 50.3 | 16.4 | 1.73 | 51.69 | 3.7 |
| 5coolV | 44.6 | −3.5 | 1.46 | 54.14 | 2.32 |
| 9 | **86.1** | 47.3 | **2.18** | 53.58 | 5.55 |
| 6 | 22.6 | −121.6 | 0.58 | 51.14 | 0.39 |

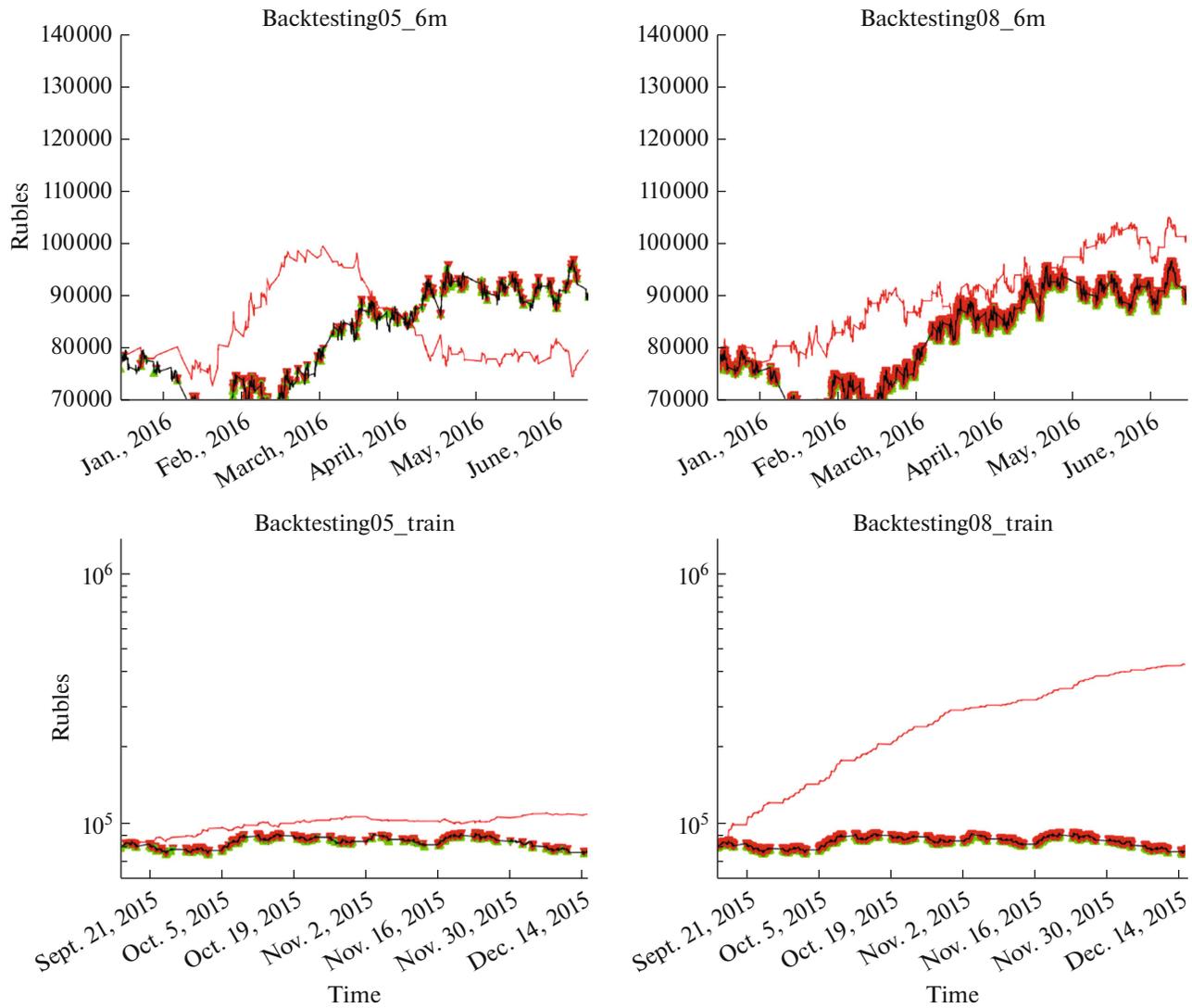

**Fig. 6.** Comparison of 5 and 8 models: effect of 64 neurons in LSTM layer (to the left) against 128 neurons. The remaining parameters are identical (Table 1).

The main difficulty in the experimental optimization of the architecture is the training time of a single model. This varies dependent on the server workload and in general counts to tens of hours. The choice of the training speed is also important for the optimization and convergence of the algorithm [17].

Below, we present the tables with the economical metrics important for the decision making on the investment attractiveness of the algorithm reflecting profitability and risk. In Table 2 we use the following denotations:

(1) Profit % per annum is the profitability in percent per annum, that is, (*profit*/*begin_price*) · (365/*number_of_days*), where *number_of_days* = 90 for three months and 180 for six months. *begin_price* is the cost of a financial instrument at the beginning of the trading.

(2) Profit % per annum (commiss.) is the profitability in percent per annum with account for the commission, that is, (*profit* − *n_trades* · *fee*/*begin_price*) · (365/*number_of_days*), where *fee* = 2.5 *rubles*.

**Table 3.** Result of execution on six-month test data

| Name | Profit % per annum | Profit % per annum (commiss.) | Sharpe ratio | Fraction of winning transactions | Average transaction, rubles |
|---|---|---|---|---|---|
| 5 | 8.8 | 5.2 | 0.29 | **55.82** | **6.15** |
| 8 | 64.0 | 28.4 | 2.14 | 52.57 | 4.5 |
| 5coolV | 75.6 | 20.7 | **2.68** | 53.24 | 3.44 |
| 9 | **110.5** | **66.5** | **3.2** | 54.21 | **6.27** |
| 6 | 8.6 | −143.7 | 0.25 | 50.84 | 0.14 |

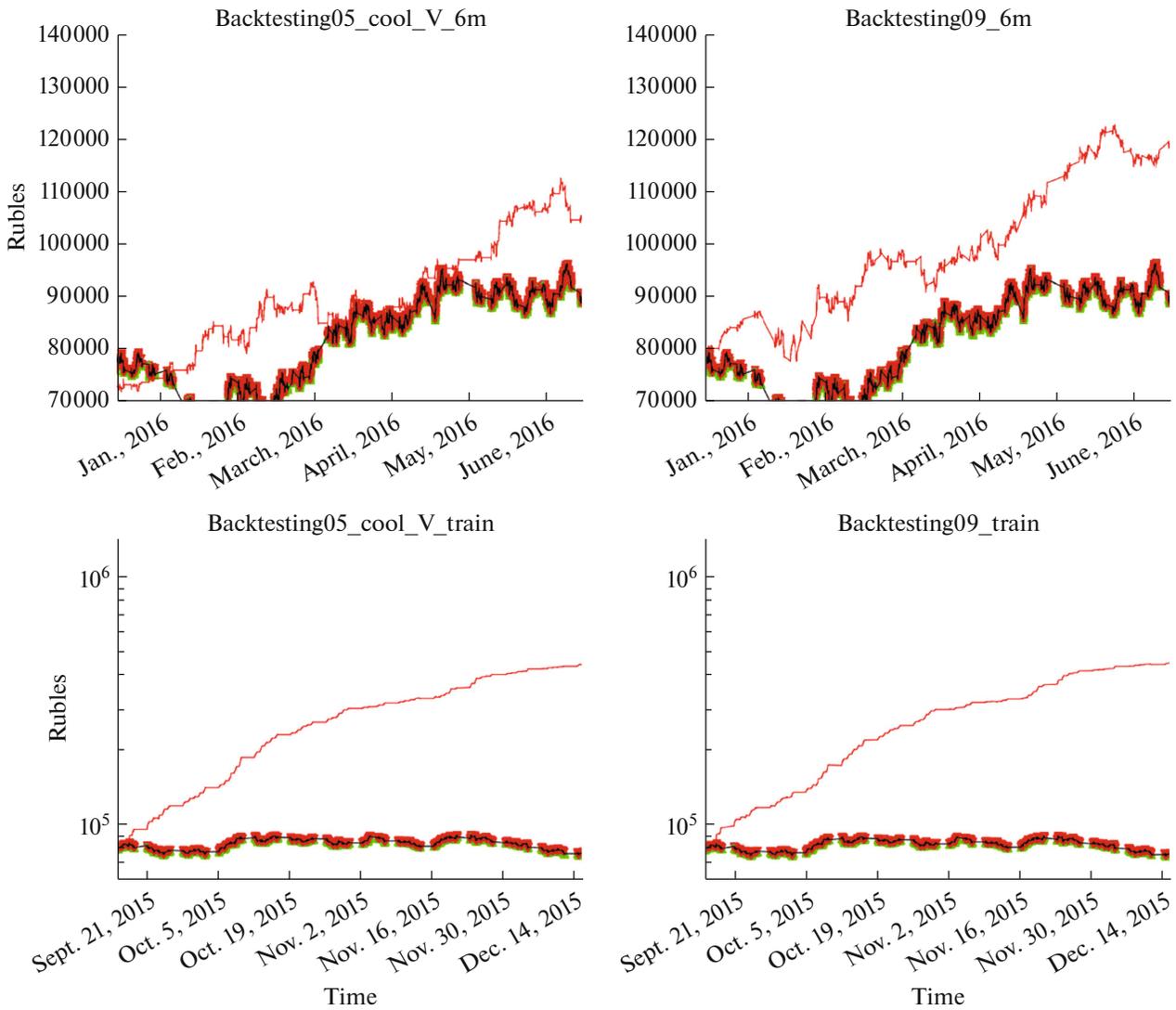

**Fig. 7.** Comparison of 5 *cool* V and 9 models: effect of combination of attributes for six steps in a common vector (to the left) against use of attributes for a single step. The remaining parameters are identical (Table 1).

**Table 4.** Result of execution on three-month training data

| Name | Profit % per annum | Profit % per annum (commiss.) | Sharpe ratio | Fraction of winning transactions | Average transaction, rubles |
|---|---|---|---|---|---|
| 5 | $10^{-3}$ | 139.1 | 5.11 | 58.13 | 47.12 |
| 8 | $5 \times 10^{-3}$ | 1784.5 | 5.11 | 68.91 | 70.28 |
| 5coolV | $10^{-3}$ | 1855.4 | 24.1 | 67.95 | 57.22 |
| 9 | $10^{-3}$ | 1894.9 | 22.57 | 70.26 | 68.21 |
| 6 | $10^{-3}$ | 7385.5 | 54.29 | 88.22 | 92.11 |

(3) Sharpe ratio is the Sharpe ratio $\mathbb{E}(profit)/\sigma(profit)$, which is the ratio of profitability to variability in both directions.

(4) Average transaction, rubles is the average profit for a transaction. This criterion is crucial in consideration of commissions.

## 5. CONCLUSIONS

This work has resulted mainly in the creation of an exchange trading algorithm based on the advantage actor-critic method, which is potentially profitable and attractive for investments from an economic point of view. Thus, the best architecture achieves a profitability of 110% per annum not accounting for commission or 66% per annum accounting for a commission of 2.5 rubles for transaction on the RTS futures (computed on six months of historical 2016 data).

During algorithm optimization, we experimentally verified several hypotheses, which allows significant improvement of method characteristics and creates a view on applicability of several ideas:

(i) The use of another reward function is disputable. On the one side, this helps avoid locking in the local minimum of absence of trading, and, on the other side, the goal function of the trader is not optimized here.

(ii) The unification of attributes for several minutes into a common vector is wrong.

(iii) The addition of a recurrent layer (LSTM) is correct.

(iv) The addition of a dropout layer is correct.

(v) The increase in the number of neurons in the hidden layers is disputable.

(vi) The use of the neuron network in several layers for approximating the cost function is correct.

As a result of the work, we also implemented a convenient environment for future experiments with exchange trading sustained in the taken style. We believe that this will allow easy experimenting using various methods to solve this problem. We think that subsequent developments of the work must be geared towards optimizing the architecture and applying it to the real trading system.

*Translated by E. Oborin*